\documentclass{PoS}
\title{
Status of indirect searches in the PAMELA and Fermi era
}
\ShortTitle{
Status of indirect searches in the PAMELA era
}
\author{\speaker{Aldo Morselli}
  \\
       INFN Roma Tor Vergata, Italy\\
       E-mail: \email{aldo.morselli@roma2.infn.it}}
\author{Igor V. Moskalenko\\
        Hansen Experimental Physics Laboratory and Kavli Institute for Particle Astrophysics and Cosmology, Stanford University, USA \\
        E-mail: \email{imos@stanford.edu}}
\abstract{The detection of $\gamma$-rays, antiprotons and positrons due
to pair annihilation of dark matter particles in the Milky Way halo
is a viable indirect technique to search for signatures of supersymmetric dark matter 
where the major challenge is the discrimination of the signal from the
background generated by standard production mechanisms.
The new PAMELA antiproton data are consistent with the standard
secondary production and this allows us to constrain exotic
contribution to the spectrum due to neutralino annihilations. In
particular, we show that in the framework of minimal supergravity
(mSUGRA), in a clumpy halo scenario (with clumpiness
factor $\geq$ 10) and for large values of $\tan(\beta)\geq 55$,
almost all the parameter space allowed by WMAP is excluded.

Instead,
the PAMELA positron fraction data exhibit an excess
that cannot be explained by secondary production. 
PPB-BETS and ATIC reported a feature in electron spectrum at a few hundred GeV.
The excesses seem to be consistent and imply
a source, conventional or exotic, of additional leptonic component.

Here we
discuss the status of 
indirect dark matter searches and 
a perspective
for
PAMELA and Fermi $\gamma$-ray space telescope (Fermi) experiments.
}

\FullConference{Identification of dark matter 2008\\
		 August 18-22, 2008\\
		 Stockholm, Sweden}
\begin{document}
\section{Antiproton to proton ratio data }
\begin{figure}[ht]
\begin{center}
\includegraphics[width=30pc]{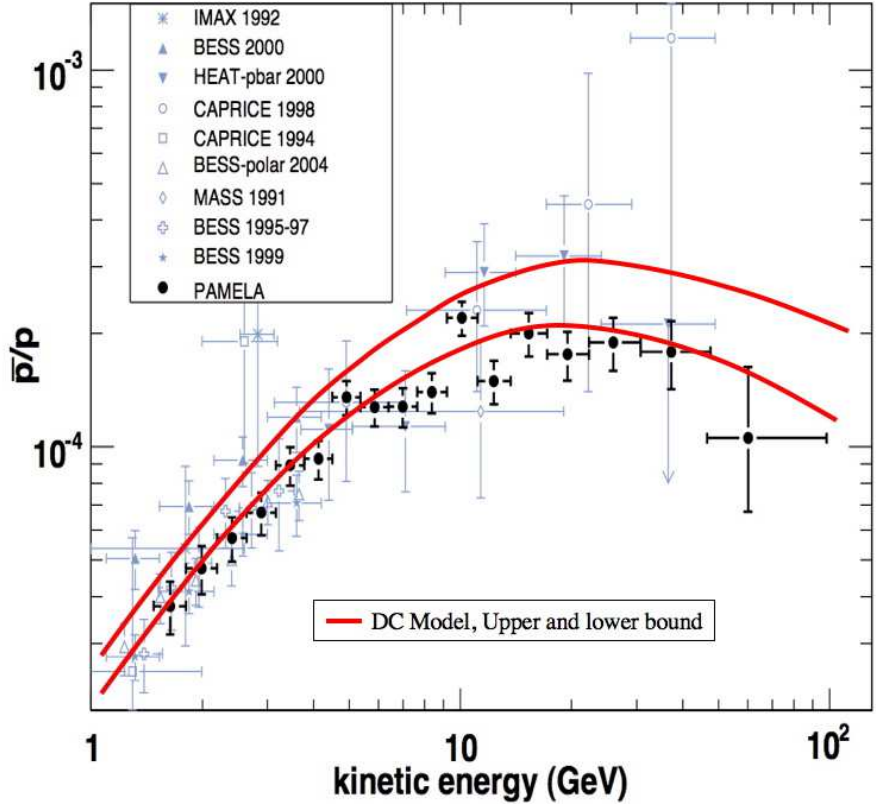}
\caption{\label{antip} 
The antiproton-to-proton flux ratio as measured by PAMELA  \cite{Pam_antip}. The lines show an approximate range expected for the standard secondary production \cite{SM134, SM2, jcap, ptuskin}.}
\end{center}
\end{figure}

\begin{figure}[ht!]

\begin{center}
\includegraphics[width=32pc]{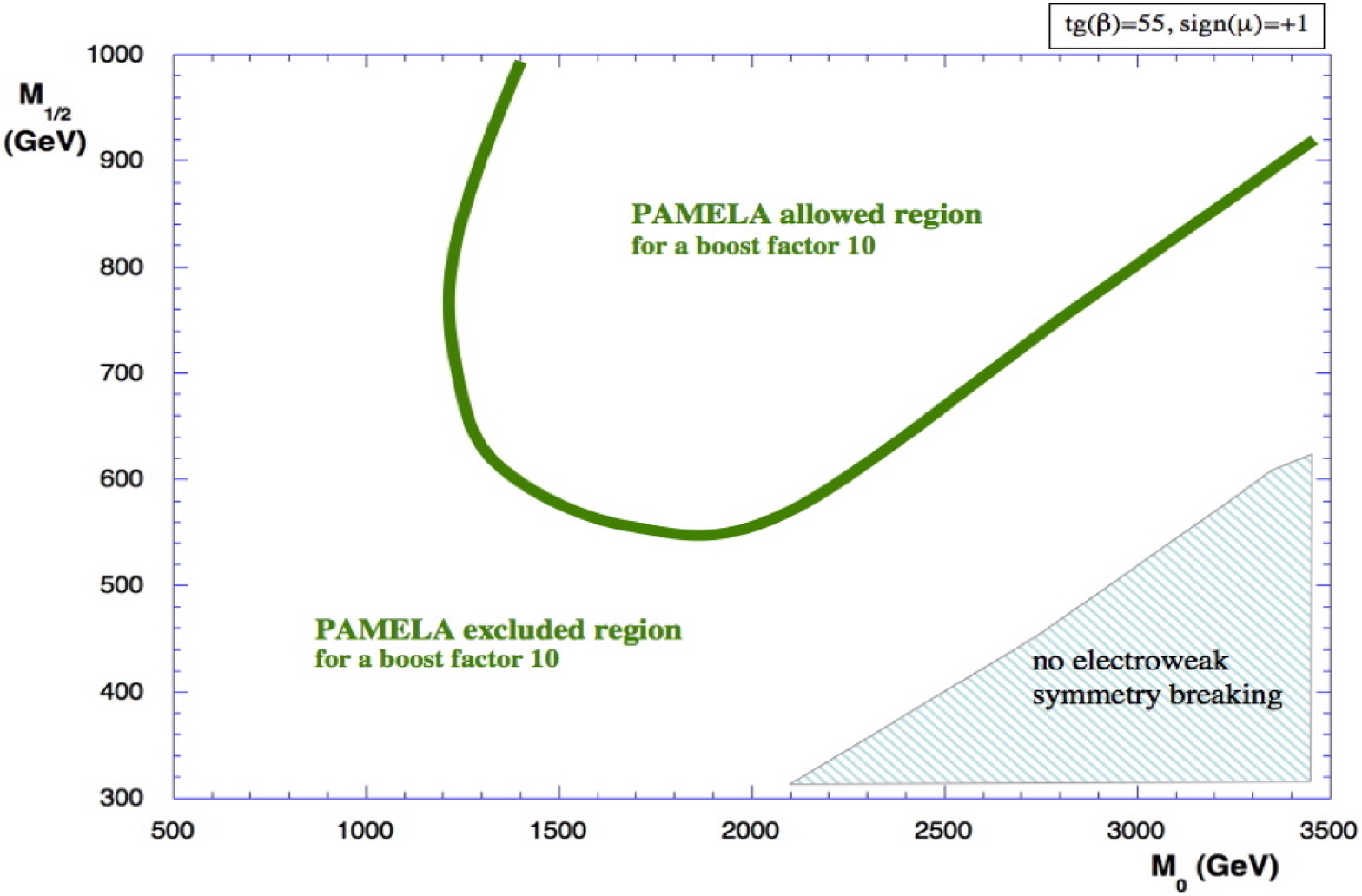}
\caption{\label{susyPAM} PAMELA excluded region in a clumpy halo scenario for a boost factor 10 in the framework of minimal supergravity  (mSUGRA) in the case of  $\tan(\beta)=55$.}
\end{center}
\begin{center}
\includegraphics[width=32pc]{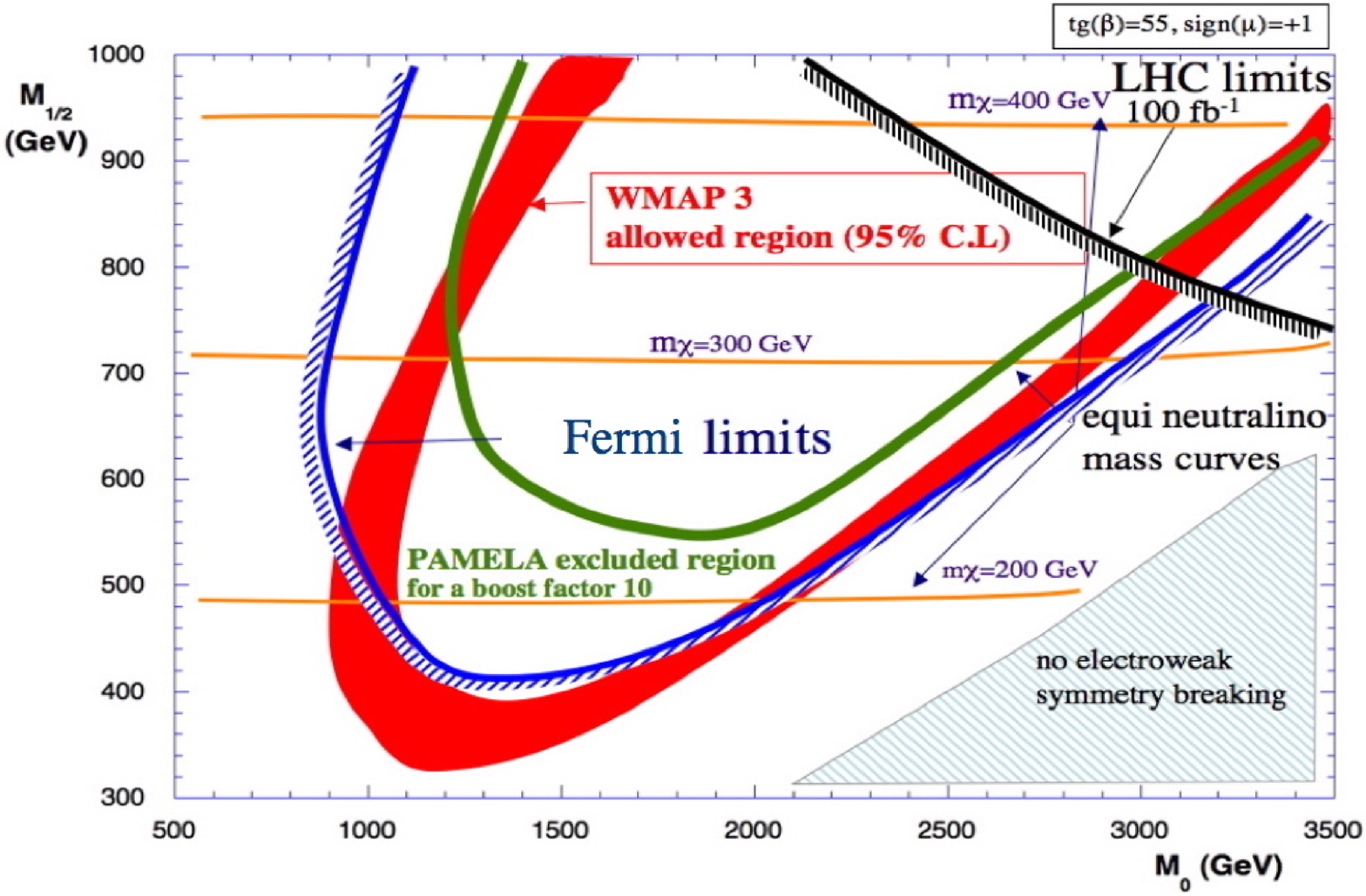}
\caption{\label{susy} Sensitivity plot for observation of mSUGRA for LHC and Fermi}
\end{center}
\end{figure}

The PAMELA (a Payload for Antimatter Matter Exploration and
Light-nuclei Astrophysics) experiment is a satellite-borne
apparatus designed to study charged particles in the cosmic
radiation with a particular focus on antiparticles (antiprotons
and positrons) \cite{PAM}. 
The PAMELA antiproton data \cite{Pam_antip} are shown in figure \ref{antip}
together with the antiproton flux expected from standard secondary production.
Cosmic ray propagation and production of secondary particles and isotopes
is calculated using the GALPROP code \cite{galprop}. 
The lines show the minimal and maximal fluxes as calculated in models with different
propagation parameters tuned to match the boron-to-carbon ratio in cosmic rays 
(\cite{SM134, SM2, jcap, ptuskin}). The antiproton data 
collected by PAMELA \cite{Pam_antip} and BESS \cite{bess} are consistent with each
other and with predictions for secondary antiproton flux thus excluding a strong 
antiproton signal from exotic processes.
Figure \ref{susyPAM} is made in the framework of minimal supergravity  (mSUGRA) by fixing the less sensitive parameters  $A_{0}$, $\tan \beta$ and ${\rm sign} (\mu) = +1$ and in the case of a clumpiness factor 10 and $\tan(\beta)=55$. 
Following the analysis in \cite{jcap},
the region below the line in figure \ref{susyPAM} can be excluded based on antiproton data. 
For larger value of tan($\beta$) the excluded parameter space is even larger, while for lower values the capability of the antiprotons flux to probe the mSUGRA scenario is very weak (\cite{jcap, ICHEP}).

This result can be compared with estimates based on Fermi five-years sensitivity 
to WIMP annihilation photons (continuum spectrum) from the Galactic center as 
shown in figure \ref{susy} \cite{dark1,dark}. 
 The red  band is the cosmologically allowed region by WMAP  \cite{WMAP};  the region above the blue line ($M_{WIMP} \sim 200$ GeV) is not observable by Fermi due to the higher WIMP mass as one moves to higher $M_{1/2}$.
The dark matter halo used for the Fermi indirect search sensitivity estimate is a truncated Navarro, Frank and White (NFW) halo profile.  For steeper halo profiles (like the Moore profile) the Fermi limits move up, covering a wider WMAP allowed region, while for less steep profile (like the isothermal profile) the Fermi limits move down, covering less WMAP allowed region.  The LHC accelerator limits are from \cite{Baer}.

\begin{figure}[ht!]
\begin{center}
\includegraphics[width=34pc]{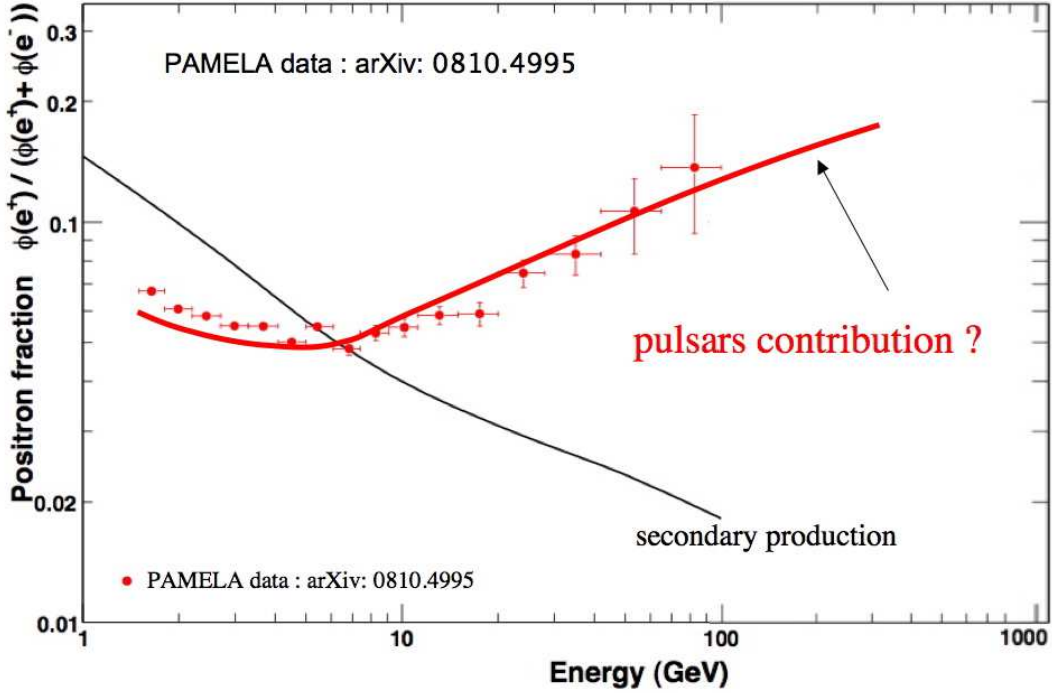}
\caption{\label{pos} PAMELA data and a possible contribution from pulsars (scaled from \cite{puls}).
Black solid line shows the background from secondary positrons in cosmic rays from GALPROP 
\cite{SM134, jcap}.}
\end{center}
\end{figure}

\section{Positron fraction and electron excess}
Contrary to the antiproton to proton ratio data, the PAMELA positron fraction data \cite{Pam_pos} exhibit  an excess above $\sim$10 GeV that cannot be explained by secondary production \cite{SM134, jcap}. 
We note that the change in the positron fraction data below $\sim$10 GeV is probably due to 
the solar modulation (e.g., \cite{clem}) and change in the polarity of the 
solar magnetic field compared to the previous cycle.
The temptation to claim the discovery of dark matter is strong, 
but  there are competing astrophysical sources, such as pulsars, that can give strong 
flux of primary positrons and electrons
\cite{puls, coutu, posAn}. In figure \ref{pos} the PAMELA data are shown with 
a possible pulsar contribution scaled from \cite{puls}.

An independent confirmation that something interesting is going on with leptons 
in cosmic rays came from measurements of high-energy electrons.
The cosmic-ray electron flux has not been measured very well in the past and especially at very-high energies
because of the very steep spectrum and thus the need for high rejection power and long exposure.
Simulations of the electron propagation from local sources \cite{kobayashi} has shown that
features in the electron spectrum may be expected in the TeV range where the flux of Galactic cosmic-ray
electrons gradually steepens. On the other hand, annihilation of Kaluza-Klein particles
may produce spectral features in sub-TeV range \cite{dark2}. 
The first indication of a feature (or excess) in the
electron spectrum at a few hundred GeV came from PPB-BETS flight a couple of years ago \cite{ppb}. 
A recent confirmation of the excess by
ATIC \cite{atic} gives more confidence that this is not an instrumental artefact. 

\begin{figure}[ht!]
\begin{center}
\includegraphics[width=36pc]{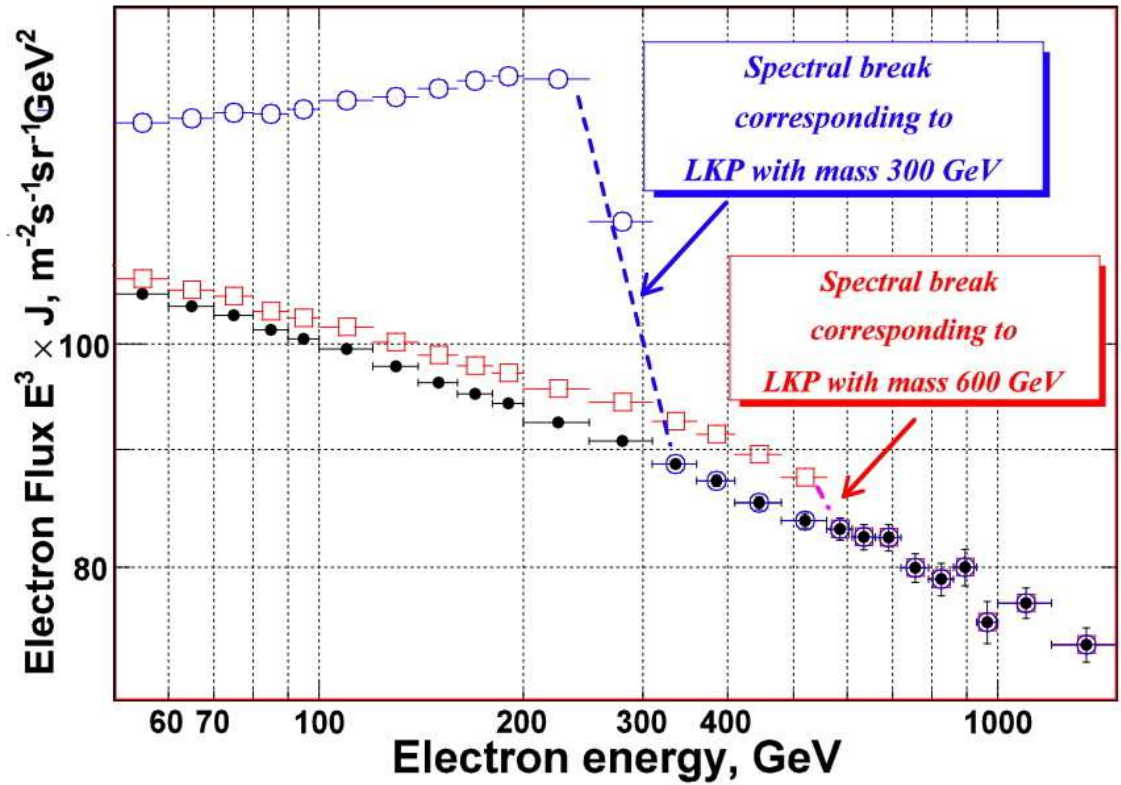}
\caption{\label{KK} 
Simulated detection of lightest  Kaluza-Klein particles (LKPs) with masses of 300 and 600 GeV in the 
LAT electron spectrum to be collected in five years of operation. Filled circles:
``conventional'' electron flux; open circles: the same but with added signal from 
300 GeV LKPs; open squares: the same with added signal from 600 GeV LKPs.
}
\end{center}
\end{figure}

How can one distinguish between the contributions of pulsars and dark matter annihilations? 
Most likely, a confirmation of the dark matter signal will require a consistency
between different experiments and new measurements of the reported excesses with large statistics.
The observed excess in the positron fraction should be consistent
with corresponding signals in absolute positron and electron fluxes in the PAMELA 
data and all lepton data collected by Fermi \cite{fermi}. 
Fermi has a large effective area and long projected lifetime, 5 years nominal with a goal 
10 years mission, which makes it an excellent detector of cosmic-ray electrons up to $\sim$1 TeV
\cite{moiseev}. 
Fermi measurements of the total lepton flux with large statistics will be able to 
distinguish a gradual change in slope with a sharp cutoff with high confidence \cite{dark2}. The latter,
as shown in figure \ref{KK}, can be an indication in favor of the dark matter hypothesis. 
A strong leptonic signal should be accompanied
by a boost in the $\gamma$-ray yield providing a distinct spectral signature
detectable by Fermi \cite{lars,nima}. Antiproton data with higher statistics and
at higher energies collected by PAMELA could also give us some clue.
%

If the sources of the excess positrons are pulsars, they should be quite close to us and, therefore, 
may be detectable in $\gamma$-rays with Fermi. In this case, one has to expect broader features in the 
electron and positron spectra without sharp cutoffs.
Meanwhile, the proposed test of the anisotropy in
the total lepton (electron+positron) flux \cite{posAn} may not work.
First, the predicted anisotropy is very small, at the fraction of a per cent.
Second, the so-called heliospheric modulation strongly affects the flux
of cosmic-ray species below 20-50 GeV. The extended heliospheric
magnetic field and the solar wind may affect the arriving directions of cosmic-ray
particles at even higher energies. Therefore, even if the anisotropy
is observed it may be connected with configuration of the heliospheric
magnetic field rather then due to the local sources of primary leptons.

\section{Conclusion}
Recent accurate measurements of cosmic-ray positrons and electrons by PAMELA, PPB-BETS, and
ATIC have open a new era in particle astrophysics. The observed features or excesses break
a boring single-power-law behavior of the cosmic-ray spectrum.
Their exotic origin has to be confirmed by complimentary findings in $\gamma$-rays by Fermi and
atmospheric Cherenkov telescopes, and by LHC in the debris of high-energy proton destructions. A positive
answer will be a major breakthrough and will change our understanding of the universe forever.
On the other hand, if it happens to be a conventional astrophysical source of cosmic rays,
it will mean a direct detection of particles
accelerated at an astronomical source, again a major breakthrough. In this case 
we will learn a whole lot about our local Galactic environment. However,
independently on the origin
of these excesses, exotic or conventional, we can expect very exciting several years ahead of us.

I.\ V.\ M.\ acknowledges support from NASA
Astronomy and Physics Research and Analysis Program (APRA) grant.

\end{document}